\pdfoutput=1
\documentclass[%
 reprint,
superscriptaddress,
 amsmath,amssymb,
 aps,
pra,
prb,
rmp,
prstab,
floatfix,
]{revtex4-2}

\usepackage{graphicx}
\usepackage{dcolumn}
\usepackage{color}
\usepackage{bm}

\begin{document}

\preprint{APS/123-QED}

\title{Investigation of the particle-particle interaction effects in the cosmic Zevatron based on cyclotron auto-resonance by particle-in-cell simulations}

\author{Yousef I. Salamin}
\email{Corresponding author: ysalamin@aus.edu}%
\affiliation{Department of Physics, American University of Sharjah, POB 26666, Sharjah, United Arab Emirates}
\author{Qian Zhao}
\email{zhaoq2019@xjtu.edu.cn}
\affiliation{
School of Physics, Xi'an Jiaotong University, Xi'an 710049, China}
\author{Ting Sun}
\email{sunting2022@stu.xjtu.edu.cn}
\affiliation{
School of Physics, Xi'an Jiaotong University, Xi'an 710049, China}

\date{\today}

\begin{abstract}
Cyclotron autoresonance acceleration has been recently advanced as a potential mechanism for accelerating nuclei to ZeV energies (1 ZeV = $10^{21}$ eV). All results have been based on single- and many-particle calculations employing analytic solutions to the relativistic equations of motion in the combined magnetic and radiation fields, excluding effects related to the particle-particle interactions. Here, results from many-particle calculations and Particle-In-Cell (PIC) simulations, are presented which lend support to the single-particle investigations. Each single-particle result is found to lie well within one standard deviation about the ensemble average obtained from the corresponding many-particle calculation. The PIC simulations show that, even for number densities far exceeding those employed in the non-interacting case, the energy gain drops markedly due to the particle-particle interactions, over the first $\sim 8~ mm$ of the acceleration length. Together with the substantial attenuation, this finding supports the conclusion that the particle-particle interaction effects can be negligibly small over acceleration lengths of typically many kilometers.
\end{abstract}

\maketitle

\section{Introduction} \label{sec:intro}
	
The binary neutron-star merger, source of the gravitational waves detected recently \cite{abbott}, was followed by the emission of ultraintense radiation, with frequencies covering a substantial part of the known electromagnetic spectrum \cite{Drout1570,Hallinan1579,Kilpatrick1583,Wu_2019}. Beamed gamma-rays, in the form of a gamma ray burst (GRB), x-rays, and visible light were emitted. Among other things, these emissions are a clear indicator of stellar nucleosynthesis and the presence of atoms \cite{pian2017spectroscopic,arcavi2017optical,smartt2017kilonova,kajino2019current,wang2020}. 
	
Investigation of the emitted radiation can, in principle, be a source of valuable information about the merger and subsequent evolution of the newly formed entity. On the other hand, interaction of charged particles with the beamed radiation, especially in the added presence of superstrong magnetic fields associated with the merging entities, can drastically influence the subsequent evolution of such particles. The question thus arises as to whether atomic nuclei can be accelerated to ZeV energies \cite{nagano,harari,Aloisio2017AccelerationAP,PhysRevD.97.063010,PhysRevLett.125.121106,PhysRevD.102.062005} and ejected as ultra-high-energy cosmic-rays (UHECR) as a result. 

This work is part of efforts dedicated to answering this question \cite{bell}. Detection of such particles is quite rare. Only 72 events, with energies exceeding 57 EeV (1 EeV = $10^{18}$ eV) were detected by the Telescope Array experiment \cite{abbasi2} between 2008 and 2013.
	
The mechanism of cyclotron autoresonance acceleration (CARA) has recently been advanced \cite{Salamin_2021} as a possible explanation for the ZeV energies of UHECR particles. Calculations have demonstrated ZeV energy gains by the nuclei of hydrogen, helium and iron, due to interaction with ultraintense radiation and superstrong uniform magnetic fields. The radiation-reaction effects were shown to be important in CARA but not to lower the energy gain substantially from the ZeV level.  
	
The investigations in \cite{Salamin_2021,salaminPLA} were general in nature and aimed at theoretical proof-of-principle demonstration of CARA \cite{1963SPhD....7..745K,PhysRevE.50.3077,PhysRevE.51.2456,PhysRevLett.76.2718,sal3,sal2,PhysRevA.62.053809,sal4,sal5} in an astrophysical context. They did not make specific reference to any known astrophysical environment where the resonance conditions (and ultraintense radiation and superstrong magnetic fields) may be found. These conditions may exist during the brief merger time of two compact objects, over a small area around the polar cap of one such object, during a magnetar-powered supernova explosion, among other possibilities  \cite{VINK2008503,kaspi,soker2017magnetar}. Away from the polar caps, topology of the steady-state magnetic field of a compact object can be much more complex than uniform and its lines can be severely curved. The requisite radiation-field intensity for CARA to work deserves some discussion, too. This is offered at the very end of Sec. \ref{sec:eqs} below.
	
As such, CARA can be put forth as a mechanism for cosmic-ray acceleration, alternative to or complementing the widely discussed models based on, for example, shock waves, magnetic reconnection and unipolar induction \cite{Aloisio2017AccelerationAP}. The existing models describe acceleration to energies close to the EeV level inside a potential cosmic-ray source, where a plasma background plays a central role. To reach the ZeV energy levels, it seems plausible to assume that a particle is first pre-accelerated inside the source by the shock wave mechanism, for example, and subsequently receives a big energy boost from CARA outside the source. This assumption will be made throughout this work.

The recent investigations employing CARA \cite{Salamin_2021} have also been single-particle. Many-particle calculations have been carried out in \cite{salaminPLA} but did not discuss the particle-particle interactions. 
In the present work, many-particle simulations are carried out to lend support to the single-particle calculations and to take into account the inter-particle corrections. Thus, the main working equations of CARA need to be amended slightly. The entities to be accelerated will be assumed to be initially picked randomly from an ensemble of $N$ particles. Shape and size of the ensemble will be decided plausibly, and $N$ will be chosen so that the number of particles per unit volume is kept way below solid density. 

The aim of this work is two-fold: (a) to support the findings of the single-particle calculations in \cite{Salamin_2021} with many-particle simulations, and (b) to strengthen the case for CARA by performing simulations which employ a more realistic set of astrophysical parameters than has been used in \cite{Salamin_2021}. Included in the latter aim is also presenting, for the first time, results for acceleration by CARA of an ensemble of nickel nuclei.
	
\begin{figure}[t]
	\includegraphics[width=8cm]{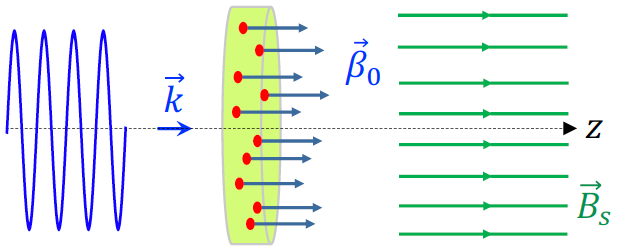}
	\caption{CARA: a many-particle schematic diagram.}
	\label{fig1}
\end{figure}

In Section \ref{sec:eqs}, the CARA working equations will be revisited in order to incorporate the set of initial conditions appropriate for an ensemble of particles. Dynamics of the ensemble of particles will be investigated, based on the revised equations, in Section \ref{sec:sim}, employing a parameter set (and for nuclei) the same as in \cite{Salamin_2021}. In section \ref{sec:real}, similar simulations will be performed for: (a) iron and nickel nuclei, the latter not covered by the single-particle calculations in \cite{Salamin_2021}, (b) a more realistic parameter set, and (c) smaller ensembles, to ensure that the particle-particle interactions may be considered negligible. Results from Particle-In-Cell (PIC) simulations, whose purpose is to shed some light on the particle-particle interaction effects, will be presented in Sec. \ref{sec:pic}. A brief discussion of our results will be conducted, and some concluding remarks will be given, in Section \ref{sec:conc}. 

\begin{figure}[t]	
\includegraphics[width=8cm]{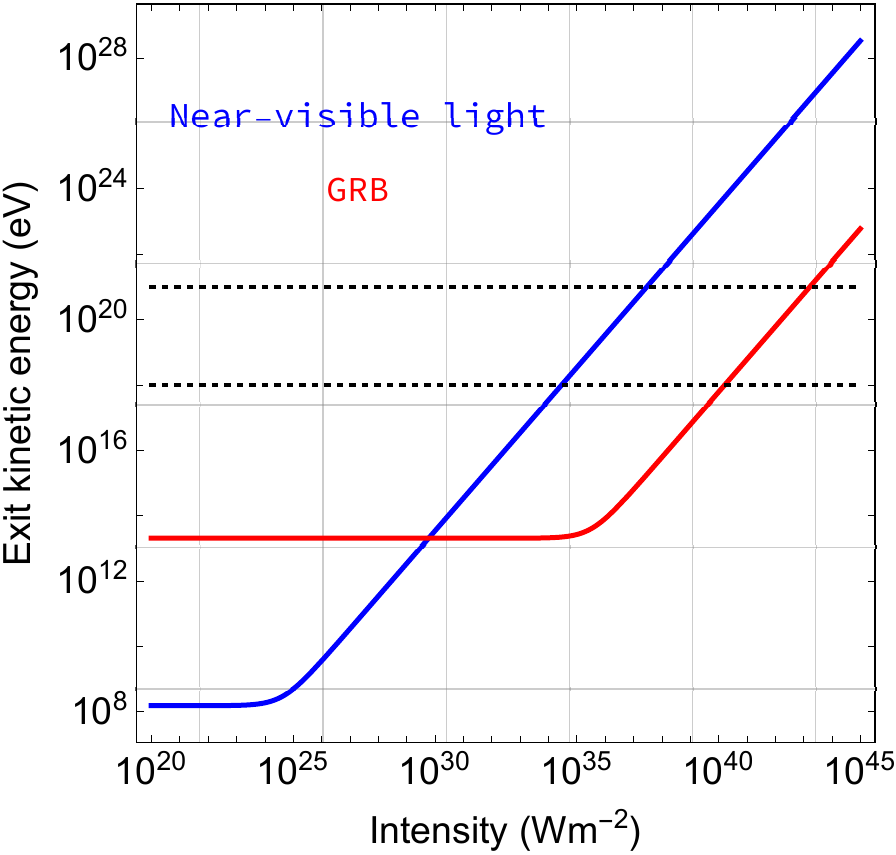}
\caption{Log-Log plot of the exit kinetic energy of the nuclide Fe$^{+26}$ with the radiation-field intensity. For acceleration by CARA using light of wavelength $\lambda=1~ \mu$m, the initial injection kinetic energy is $K_0 = 150$ MeV. For the GBR of wavelength $\lambda=5\times10^{-11}$ m, $K_0=20$ TeV. In both plots interaction is with 5 radiation-field phase-cycles. The horizontal dotted lines represent exit kinetic energies of 1 EeV and 1 ZeV, respectively. }
\label{fig2}
\end{figure}

\section{The equations}\label{sec:eqs}
	
Figure \ref{fig1} is a schematic diagram showing the initial ensemble of $N$ identical particles, each of mass $M$ and charge $Q$, moving along the $z$-axis of a Cartesian coordinate system. Their positions are uniformly distributed inside a cylinder of radius $R$ and height $H$. Their initial speeds are derived from a normal distribution of their injection kinetic energies, of mean $\bar{K}_0$ and standard deviation $\Delta K_0$. This makes the number density $n_d=N/(\pi R^2H)$. The schematic diagram also shows a uniform magnetic field of strength $B_s$, oriented along $+z$, and a radiation wave propagating along the same direction. Only the size and shape of the initial ensemble, and the mean and spread of the initial kinetic energies, will be fixed. In all many-particle calculations in this work, the initial ensemble size will be determined by the choices $R=5$ m and $H=\lambda$, the wavelength of the radiation field employed. 

The electromagnetic fields will be modeled by \cite{Salamin_2021}
\begin{eqnarray}
\label{E}\bm{E} &=& \hat{\bm{i}} E_0 \sin\eta,\\
\label{B}\bm{B} &=& \hat{\bm{j}} \frac{E_0}{c} \sin\eta+\hat{\bm{k}} B_s.
\end{eqnarray}
In these equations, $E_0$ is the constant amplitude of the plane-wave radiation, $c$ is the speed of light in vacuum, $\eta = \omega t-kz$, $k = \omega/c$, and $\hat{\bm{i}}, \hat{\bm{j}}$ and $\hat{\bm{k}}$ are unit vectors in the positive $x$-, $y$- and $z$-directions, respectively. Recall, at this point, the resonance condition that characterizes CARA. It ties the particle and radiation and magnetic field parameters by $r=1$, where
\begin{equation}\label{r}
	r=\frac{\omega_c}{\omega}\sqrt{\frac{1+\beta_0}{1-\beta_0}};\quad \omega_c=\frac{QB_s}{M},
\end{equation}
	in which $\omega$ is the radiation frequency and $\omega_c$ is the cyclotron frequency of the particle around the lines of the magnetic field. Resonance occurs when the cyclotron frequency matches the Doppler-shifted frequency of the radiation field, which the particle senses in its own rest frame. With $Q, M$, and $\omega$ fixed, the resonance condition is essentially a relationship between $B_s$ and $\beta_0$, the randomly-selected initial speed.
	
\begin{table*}
\caption{\label{tab:tab1}
	Parameters, exit kinetic energies, and resonance magnetic fields for $N=100$ protons accelerated by CARA.}
\begin{ruledtabular}
\begin{tabular}{cccc}
\textrm{Intensity (W/m$^2$)} &
\textrm{$\lambda$ (m)} &
\textrm{$\bar{K}_e\pm\Delta K_e$ (ZeV)} &\textrm{$\bar{B}_s\pm\Delta B_s$ (MT)}\\
\hline
$10^{38}$ & $10^{-6}$ & $0.437\pm0.002$  & $11.25\pm0.03$  \\
$10^{43}$ & $5\times10^{-11}$ & $2.663\pm 0.028$  & $9.22\pm0.09$  \\
\end{tabular}
\end{ruledtabular}
\end{table*}

We proceed now to amend the solutions to the equations of motion of a single particle in the presence of the electromagnetic fields, by properly incorporating the above-mentioned initial conditions. The obtained equations will be used in the next section to carry out the promised many-particle simulations. On-resonance solutions to the relativistic Newton-Lorentz equations of motion follow essentially the same steps as in \cite{Salamin_2021}. With the initial conditions on position expressed as $\eta_0=-kz_0$, one finally obtains  
\begin{widetext}
\begin{eqnarray}\label{xres}
	x(\eta) &=& x_0+\frac{ca_0}{2\omega}\gamma_0(1+\beta_0) \left[(\sin\eta-\sin\eta_0)-(\eta\cos\eta-\eta_0\cos\eta_0)\right],\\
\label{yres}
	y(\eta) &=& y_0+\frac{ca_0}{2\omega}\gamma_0(1+\beta_0) \left[(\eta\sin\eta-\eta_0\sin\eta_0)+2(\cos\eta-\cos\eta_0)\right],\\
	\label{zres}
z(\eta) &=& z_0+\frac{c}{\omega}\left(\frac{1+\beta_0}{1-\beta_0}\right) \left\{\left(\frac{\beta_0}{1+\beta_0}\right)(\eta-\eta_0)+\left(\frac{a_0^2}{24}\right)(\eta-\eta_0)^2(\eta+2\eta_0)\right.\nonumber\\
& &\left.+\frac{a_0^2}{16}\left[(\eta-2\eta_0) \cos2\eta_0+\eta\cos2\eta+(2\eta_0^2-2\eta\eta_0-1)\sin2\eta_0-\sin2\eta\right]\right\},\\
\label{gammares}
	\gamma(\eta) &=& \gamma_0+\frac{a_0^2}{8}\gamma_0(1+\beta_0)\left[(\eta^2-\eta_0^2)+(\sin^2\eta-\sin^2\eta_0)-(\eta\sin2\eta-\eta_0\sin2\eta_0)\right].
\end{eqnarray}
\end{widetext}
In these equations, $\gamma_0=(1-\beta_0^2)^{-1/2}$ and $a_0\equiv QE_0/(M\omega c)$. Note that $a_0^2$ may be thought of as a dimensionless radiation-field intensity parameter, whereas the radiation field intensity in W/m$^2$ is $I=c\epsilon_0E_0^2/2$, where $\epsilon_0$ is the permittivity of free space. Equations (\ref{xres})-(\ref{zres}) give a parametric representation of the particle's trajectory. Equation (\ref{gammares}) is the particle's Lorentz factor (its energy scaled by $Mc^2$). 

For the special case of initial position at the origin of coordinates, evolution of the kinetic energy of a particle with $\eta$ may be written as \cite{Salamin_2021}
\begin{eqnarray}
K(\eta) &=& K_0+ \left[\frac{Q^2}{16\pi^2\epsilon_0 Mc^3}\right]\gamma_0(1+\beta_0) (I\lambda^2)\nonumber\\
& &\hskip2cm\times\left[\eta^2+\sin ^2\eta-\eta\sin2\eta\right],
\end{eqnarray}\label{K1} 
where $K_0=(\gamma_0-1)Mc^2$ is the initial injection kinetic energy. Figure \ref{fig2} shows log-log plots of the exit kinetic energies against $I$, the radiation-field intensity, at the end of interaction with 5 phase cycles of near-visible light and a GRB. The main assumption here is that the particle is pre-accelerated to kinetic energies of 150 MeV (visible) and 20 TeV (GRB). For these initial conditions, the resonance magnetic field strengths are 38.9316 MT and 1.09132 GT, respectively.  Note that the flat parts of the $K_e$ vs. $I$ curves reflect those initial injection energies. The particle's kinetic energy begins to increase substantially after some threshold intensity has roughly been passed ($\sim 10^{25}$ W/m$^2$, for near-visible light, and $\sim 10^{35}$ W/m$^2$, for the GRB). The energy range of 1 EeV to 1 ZeV is bounded by the two horizontal dotted lines in Fig. \ref{fig2}. The figure clearly shows that, for the chosen parameters, the minimum radiation-field intensities required to reach the EeV to ZeV kinetic energy levels are $\sim10^{35}$ W/m$^2$ (near-visible) and $\sim10^{42}$ W/m$^2$ (GRB).

\begin{figure*}[htb]
	\includegraphics[width=7cm]{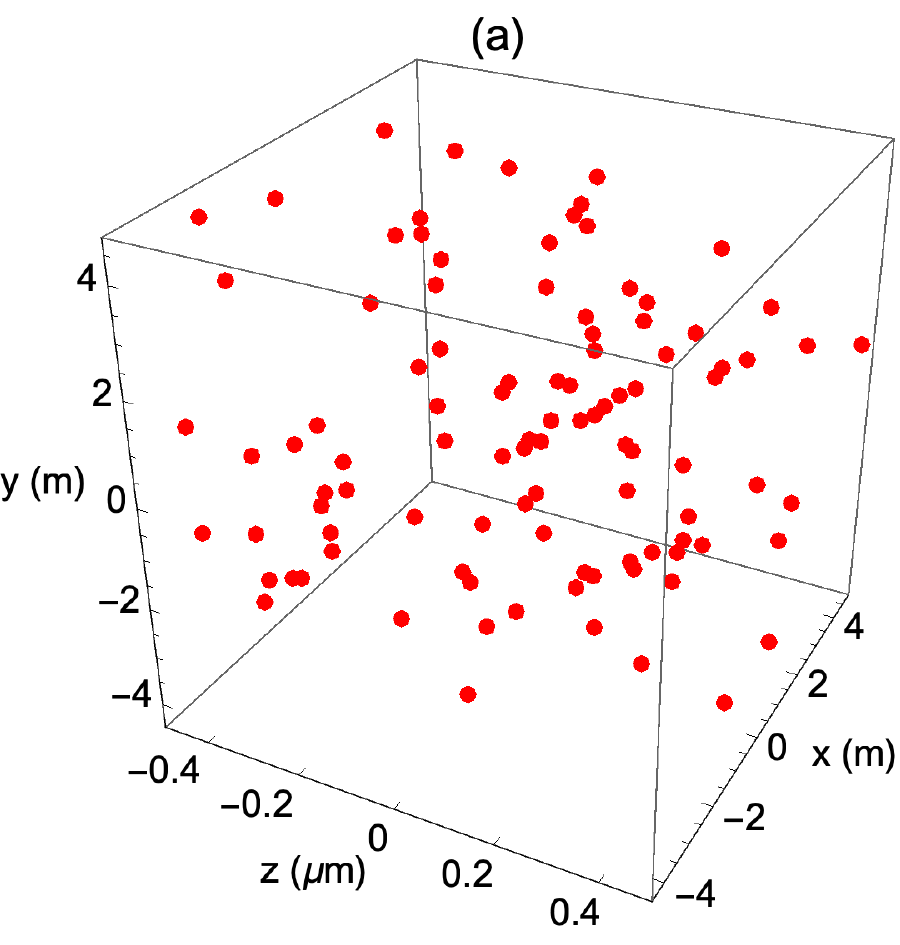}
	\includegraphics[width=7cm]{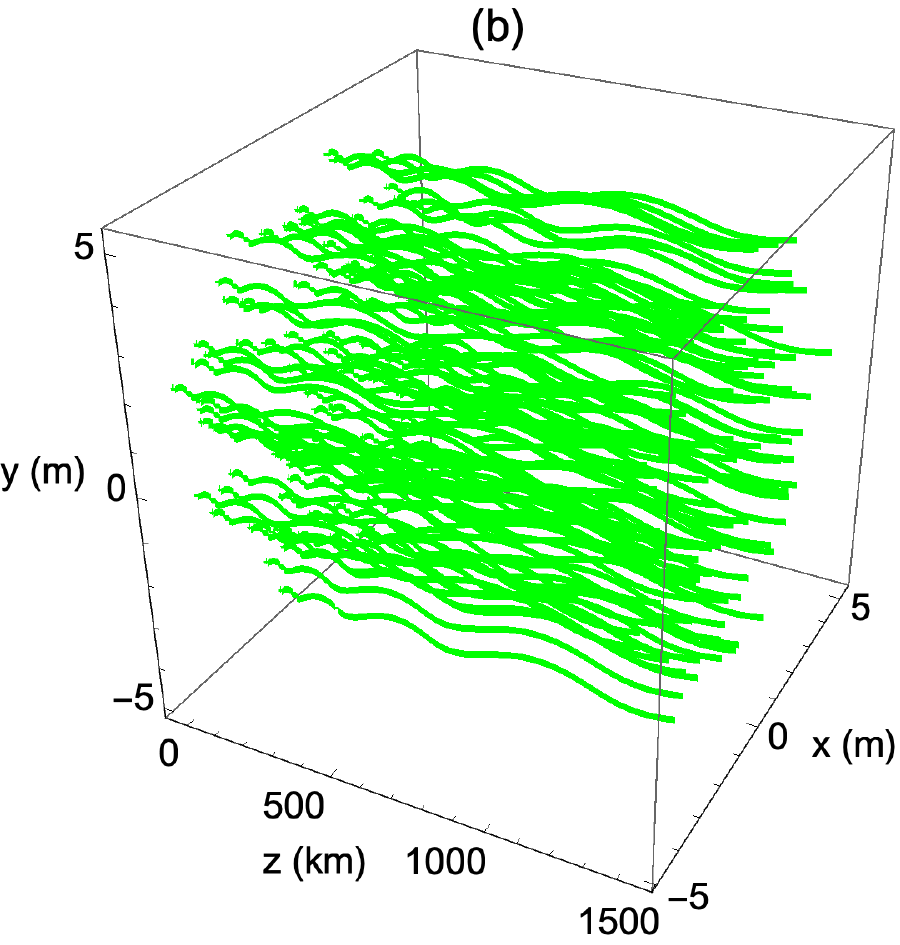}
	\includegraphics[width=7cm]{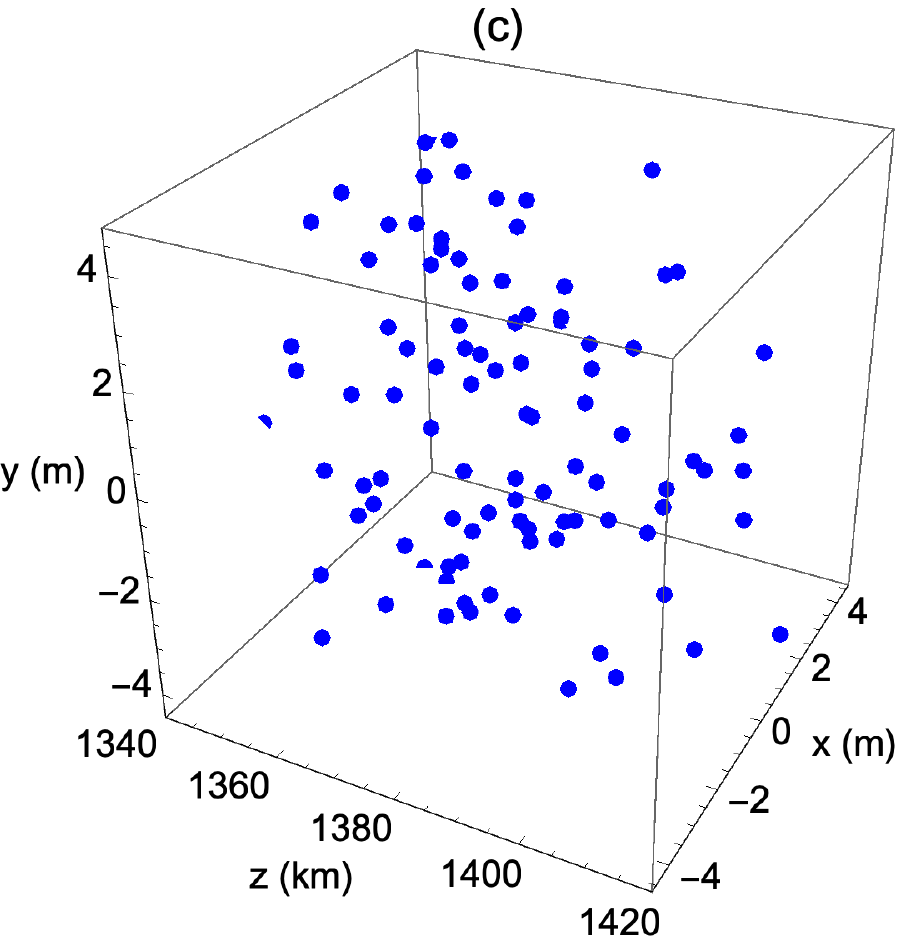}
	\includegraphics[width=7cm]{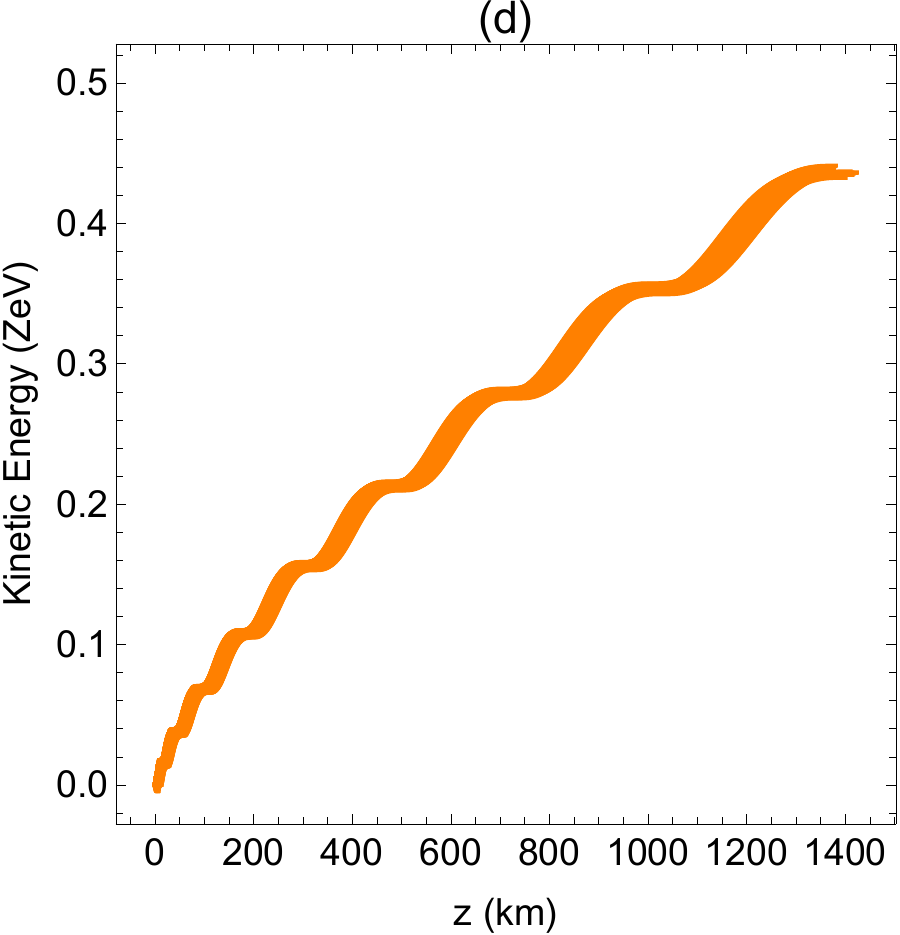}
	\caption{Proton acceleration by CARA employing near-visible light. (a) Initial ensemble of $N = 100$ protons inside a disk of radius 5 m and thickness $\lambda = 1~\mu$m (number density $n_d \simeq 1.27324\times10^6$ m$^{-3}$). Initial ensemble kinetic energy: normal distribution of mean $\bar{K}_i=150$ MeV and standard deviation $\Delta K_0=1.5$ MeV. (b) Actual trajectories of the ensemble members during interaction with the radiation and magnetic fields. (c) Distribution of the ensemble particles at the end of an interaction time equivalent to 5 phase cycles ($\Delta\eta=10\pi$) of the radiation field. (d) Kinetic energy evolution with the excursion distance for all of the particles in the ensemble. The radiation field intensity is $I=10^{38}$~ W/m$^2$ and the resonance magnetic field strength sensed by the ensemble members and calculated on the basis of Eq. (\ref{r}) is $B_s = 11.2529\pm0.0354$ MT.}
	\label{fig3}
\end{figure*}

\section{The many-particle calculations}\label{sec:sim}
	
The particle ensemble dynamics in the combined radiation plus uniform magnetic fields will now be discussed using the equations in terms of the radiation field phase. Without loss of generality, the examples will focus on the nuclei H$^{+1}$, He$^{+2}$ and Fe$^{+26}$, as in \cite{Salamin_2021}. Cosmic rays are close to 90\% protons, H$^{+1}$, the simplest atomic nucleus. Alpha particles, He$^{+2}$, account for about 10\%, and the rest are heavier nuclei. Fe$^{+26}$ is one of the most stable nuclei in nature. Furthermore, recent measurements by the Pierre Auger Observatory \cite{Aab} suggest that most UHECR particles are nuclei heavier than the proton. 
	
\begin{figure*}[htb]
\includegraphics[width=7cm]{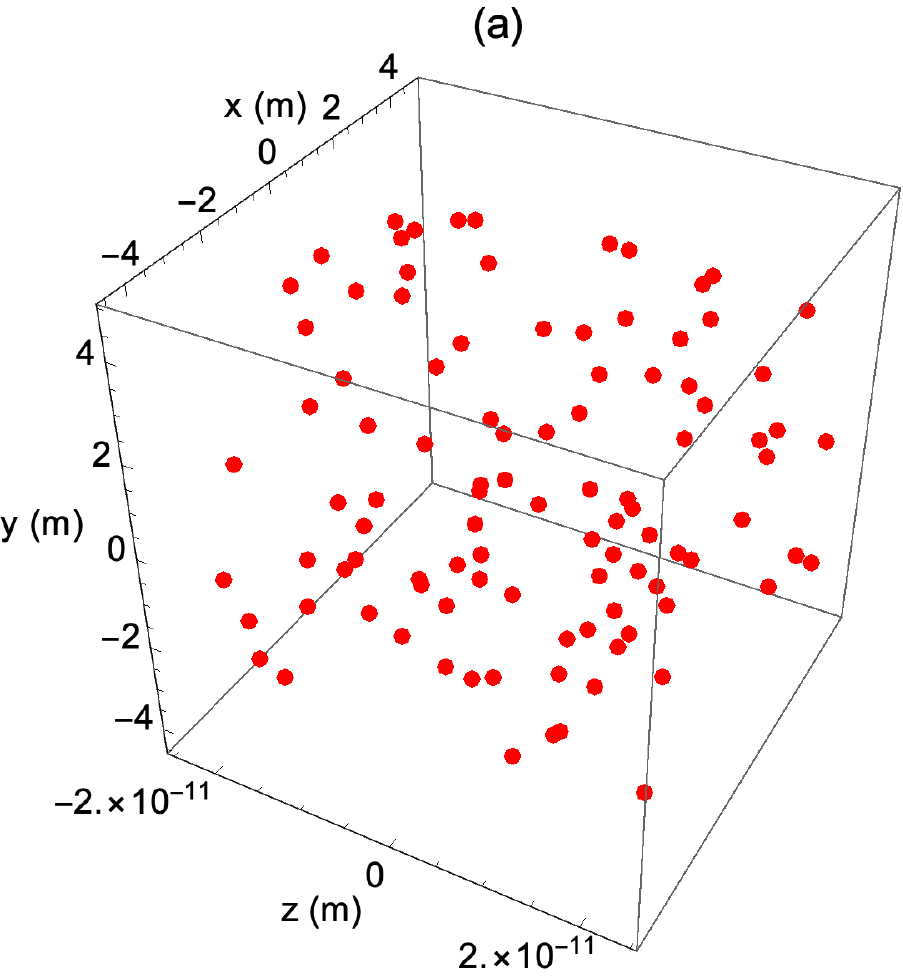}
\includegraphics[width=7.cm]{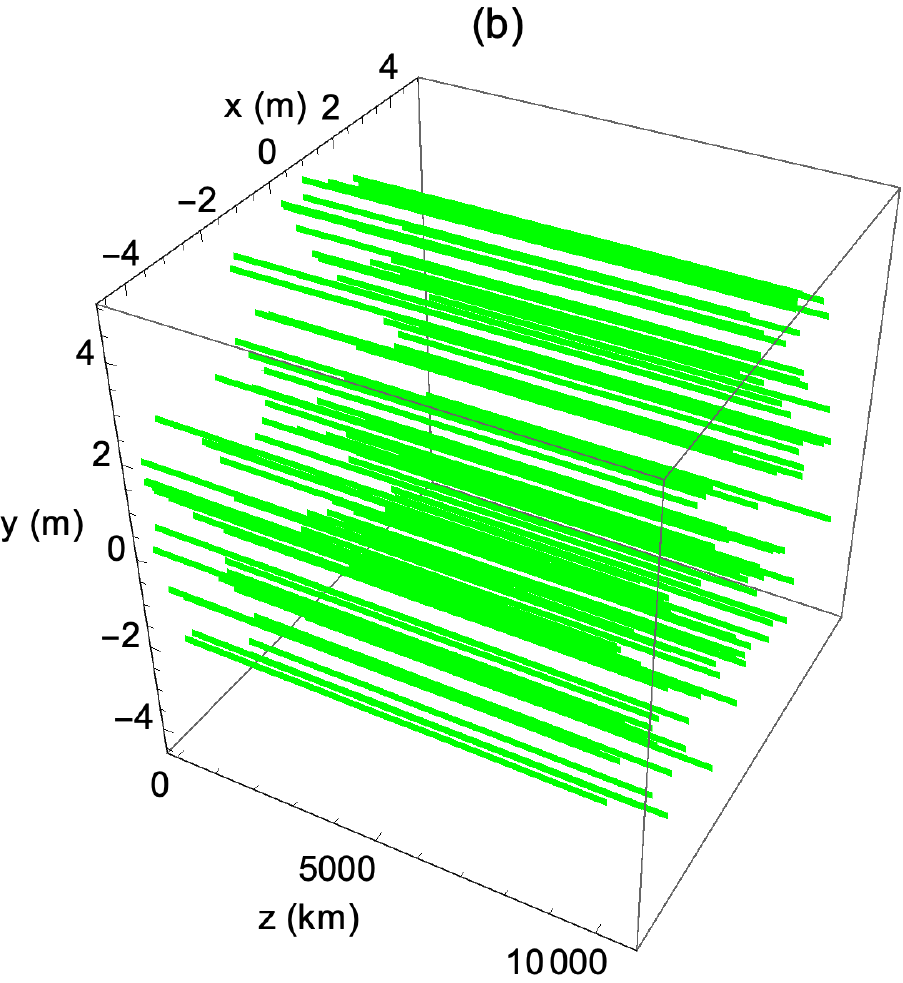}
\includegraphics[width=7.cm]{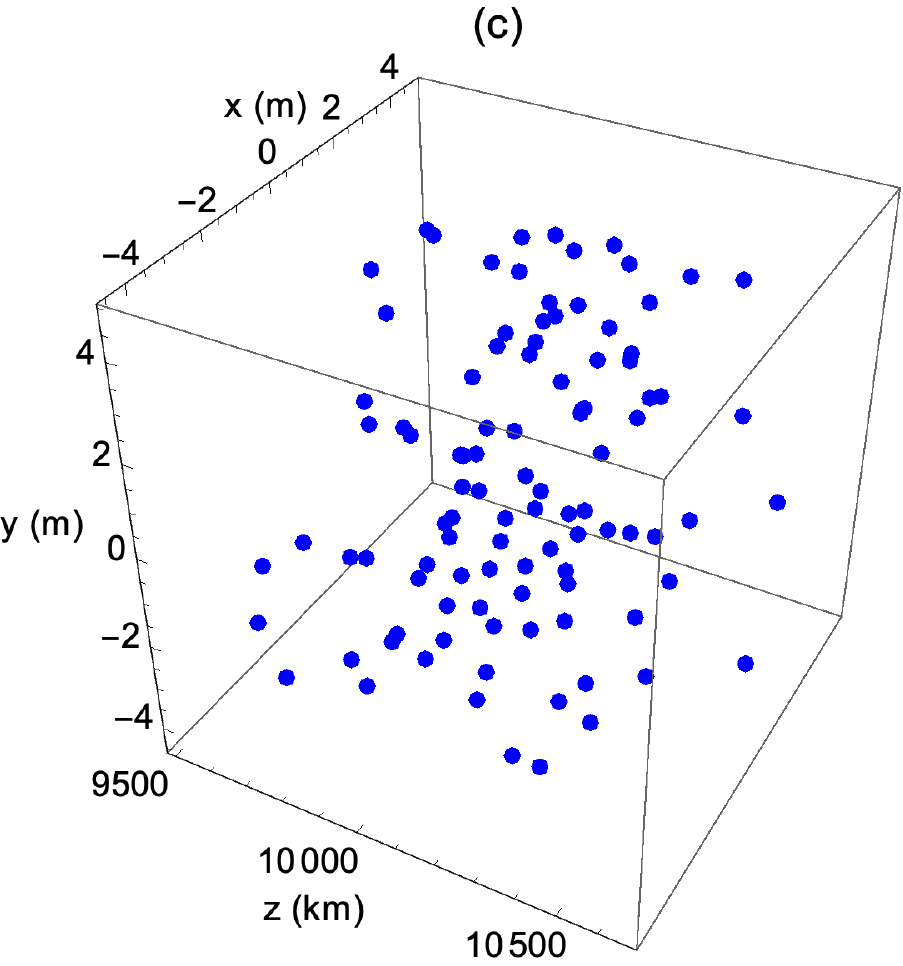}
\includegraphics[width=7.cm]{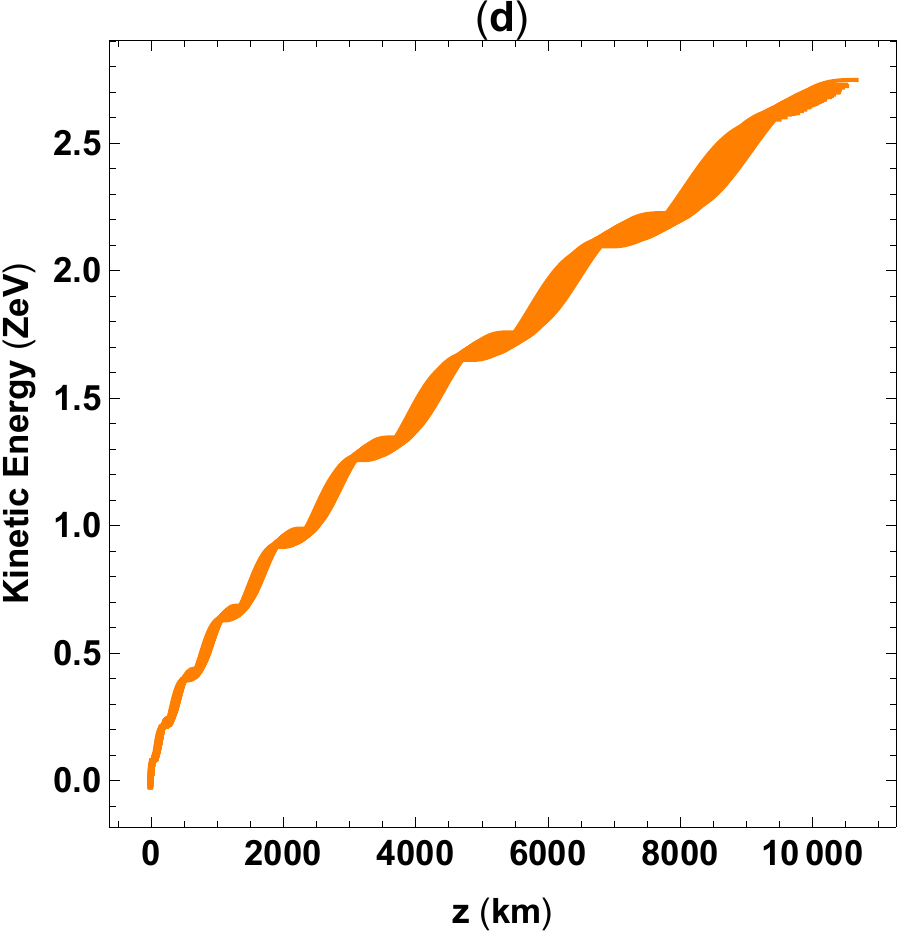}
\caption{Acceleration by CARA of protons employing the fields of a GRB. (a) Initial ensemble of $N = 100$ protons inside a disk of radius 5 m and thickness $\lambda = 5\times10^{-11}$ m (number density $n_d \simeq 2.54648\times10^{10}$ m$^{-3}$). Initial ensemble kinetic energy: normal distribution of mean $\bar{K}_0=20$ TeV and standard deviation $\Delta K_0=0.2$ TeV. (b) Actual trajectories of the ensemble members during interaction with the radiation and magnetic fields. (c) Distribution of the ensemble particles at the end of an interaction time equivalent to 5 phase cycles ($\Delta\eta=10\pi$) of the radiation field. (d) Kinetic energy evolution with the excursion distance for all of the particles in the ensemble. The radiation field intensity is $I=10^{43}$~ W/m$^2$ and the resonance magnetic field strength sensed by the ensemble members and calculated on the basis of Eq. (\ref{r}) is $B_s = 9.23921\pm0.09534$ MT.}
\label{fig4}
\end{figure*}

As shown schematically in Fig. \ref{fig1}, members of the initial ensemble are assumed to have been pre-accelerated to relativistic velocities along the common directions of $\bm{B}_s$ and $\bm{k}$, the latter being the wavevector of the radiation field, by shock waves or any other means \cite{schlickeiser}. For simplicity, it will be assumed that the wavefront of a radiation wave catches up with a particle at $t=0$ when the latter is at the spatial coordinates ($x_0, y_0, z_0$) and has a speed $\beta_0$, derived from the corresponding initial normal distribution of kinetic energies of mean $\bar{K}_0$ and standard deviation $\Delta K_0$. In all cases considered, the interaction time will be equivalent to five radiation-field phase cycles, $\Delta\eta=\eta_e-\eta_0=10\pi$, with $e$ standing for {\it exit}. 

As examples, we first investigate the dynamics of $N=100$ protons accelerated by CARA, without reference to any astrophysical environment, known to a good degree of certainty, where the  conditions for acceleration may be met \cite{hillas,drury,chen,aharonian3,honda,drury2,osmanov,fang,Liu_2017}. Figure \ref{fig3} displays the results for acceleration using near-visible light of intensity $I=10^{38}$ W/m$^2$ and wavelength $\lambda=1~\mu$m. Speeds of the particles of the initial ensemble are derived from a normal distribution of kinetic energies ($\bar{K}_0=150$ MeV, and $\Delta K_0=1.5$ MeV). Figure \ref{fig3}(a) shows the initial ensemble, a uniform distribution of initial positions ($x_0, y_0, z_0$). Interactions are assumed to commence at $t=0$ (or, equivalently, at $\eta_0=-kz_0$) causing the particles to follow the trajectories shown in Fig. \ref{fig3}(b). Figure \ref{fig3}(c) shows the positions through which the particles pass at the end of the interaction time. In other words, the initial ensemble in (a) evolves to the final spatial distribution of particles shown in (c) as a result of the acceleration process. Assuming that each particle's own initial conditions launch it into cyclotron autoresonance, not necessarily exactly, this will result in tremendous energy gain. In Fig. \ref{fig3}(d) the exit kinetic energy of each particle of the ensemble is shown as a function of its axial excursion along the $z$-direction. Exit (end-of-interaction) results for this example are displayed in the first row of Table \ref{tab:tab1}. Note, in particular, that the magnetic field strength shown in the last column is given as a mean $\pm$ some spread. This is due to the fact that once a value for $\beta_0$ has been picked at random, a value for $B_s$ will be dictated by the resonance condition, Eq. (\ref{r}). Nevertheless, the spread in those values does not seem to disturb resonance appreciably and the particles end up attaining ZeV kinetic energies.
	
\begin{table*}
\caption{\label{tab:tab2}%
		Exit statistics of $10^4$ particles accelerated by CARA, employing ultraintense near-visible light. }
\begin{ruledtabular}
\begin{tabular}{cccccc}
		\textrm{Nucleus} &
		\textrm{$\bar{K}_e\pm\Delta K_e$ (ZeV)} &
		\textrm{$\bar{x}_e\pm\Delta x_e$ (m)} &
		\textrm{$\bar{y}_e\pm\Delta y_e$ (m)} &
		\textrm{$\bar{z}_e\pm\Delta z_e$ (km)} &
		\textrm{$\bar{B}_s\pm\Delta B_s$ (MT)} \\
		\hline
		H$^{+1}$ &  $0.437\pm0.002$ &  $-0.22\pm2.48$ &  $-0.001\pm2.516$ &  $1352\pm12$ &  $11.25\pm0.03$  \\
		He$^{+2}$ &   $0.334\pm0.001$ &   $-0.05\pm2.49$ &  $-0.029\pm2.487$ &  $197.5\pm1.4$ &   $29.44\pm0.04$  \\
		Fe$^{+26}$ &   $3.281\pm0.013$ &  $-0.06\pm2.49$ &  $0.011\pm 2.503$ & $113.0\pm0.7$ & $38.93\pm0.01$ \\
	\end{tabular}
\end{ruledtabular}
\end{table*}

The second illustrative example also involves acceleration of $N=100$ protons, albeit employing the fields of a GRB of intensity $I=10^{43}$ W/m$^2$ and wavelength $\lambda=5\times10^{-11}$ m. In this case, the initial ensemble kinetic energy (normal) distribution has mean $\bar{K}_0=20$ TeV and spread $\Delta K_0=0.2$ TeV \cite{kann2019highly,paolo}. Figure \ref{fig4} displays results of simulations for this example similar to those of Fig. \ref{fig3}. Numerical values of the exit dynamical quantities pertaining to this example are displayed in the second row of Table \ref{tab:tab1}. The exit kinetic energies of the protons from interaction with the GRB are substantially larger than from interaction with the lower-intensity near-visible radiation, as expected. The resonance magnetic field in this case is lower than in the case of interaction with the lower-frequency near-visible light, as Eq. (\ref{r}) predicts.
	
Further results from simulations performed essentially along the same lines, albeit involving a much bigger ensemble, two more nuclei, and exhibiting more exit numerical values, will be presented next. Only the exit numerical values will be displayed in tabular format, for the nuclei H$^{+1}$, He$^{+2}$, and Fe$^{+26}$. Table \ref{tab:tab2} shows results for the acceleration of $N=10^4$ particles by CARA, employing near-visible light of intensity $I=10^{38}$ W/cm$^2$ and wavelength $\lambda=1~\mu$m. Fairly good estimates of the exit mean and spread of the kinetic energy and spatial coordinates, as well as the resonance magnetic field strength, may be read from the tabulated results. For example, from the last row for iron, one concludes that an ensemble of $N=10^4$ nuclei uniformly distributed initially inside a cylinder of radius $R=5$ m and height $H=\lambda=1~\mu$m, evolves into roughly a cylinder bounded by a box of dimensions $2\Delta x_e\sim5$ m, $2\Delta y_e\sim5$ m, and $2\Delta z_e\sim1.4$ km. 
		
\begin{table*}[htb]
\caption{\label{tab:tab3}Exit statistics of $10^4$ particles accelerated by CARA, employing parameters of a GRB.}
\begin{ruledtabular}
\begin{tabular}{cccccc}
		\textrm{Nucleus} & \textrm{$\bar{K}_e\pm\Delta K_e$ (ZeV)} &
		\textrm{$\bar{x}_e\pm\Delta x_e$ (m)} &
		\textrm{$\bar{y}_e\pm\Delta y_e$ (m)} &
		\textrm{$\bar{z}_e\pm\Delta z_e$ (km)} &
		\textrm{$\bar{B}_s\pm\Delta B_s$ (MT)} \\
  \hline
		H$^{+1}$ & $2.664\pm0.029$ & $-0.030\pm2.507$ & $0.003\pm2.486$ & $10064\pm213$ & $9.22\pm0.09$  \\
		He$^{+2}$ & $0.675\pm0.007$ & $-0.021\pm2.501$ & $0.044\pm2.500$ & $161.6\pm3.4$ & $72.8\pm0.7$  \\
		Fe$^{+26}$ & $0.585\pm0.006$ & $0.024\pm2.497$ & $0.014\pm2.509$ & $0.72\pm0.02$ & $1091\pm11$  \\
\end{tabular}
\end{ruledtabular}
\end{table*}
	
Table \ref{tab:tab3} is similar to Table \ref{tab:tab2}, but using the fields of a GRB of intensity $I=10^{43}$ W/m$^2$ and wavelength $\lambda=5\times10^{-11}~$m. The results presented in Tables \ref{tab:tab2} and \ref{tab:tab3} follow different patterns, as they correspond to two widely differing sets of parameters. Their injection energies, radiation field intensities, and radiation wavelengths differ by about 5 orders of magnitude. This leads to different resonance magnetic field strengths. In Table \ref{tab:tab3}, $K_e$ decreases with increasing mass, due to its dependence on the ratio $Z^2/A$, where $Z$ is the atomic number and $A$ is the mass number of the particle \cite{Salamin_2021}. In Table \ref{tab:tab2} $K_e$ follows the opposite trend, due to increasing resonance magnetic field values, in jumps of one order of magnitude. The effect on $K_e$ in Table \ref{tab:tab3} due to $B_s$ is not great because the $B_s$ values there are comparable.

\section{A more realistic parameter space}\label{sec:real}

It may be argued that the parameters employed in our calculations thus far have been unrealistic. The assumption has been made that a big portion of the energy output from the source, like a binary-star merger, is radiant and beamed through a small circle, which leads to the GRB intensities exceeding $10^{42}$ W/m$^2$, for example, that have been employed in \cite{Salamin_2021}. In the scientific literature of relevance, the assumption is often made that energy is radiated isotropically, not in a beam. The intensity calculated based on this assumption must, therefore, be many orders of magnitude smaller than $10^{42}$ W/m$^2$. 

Another assumption made in our many-particle calculations has been that the particle-particle interactions are negligible. This has been justified by the fact that the number densities employed are very small compared to those in a typical solid, where such interactions can not be ignored. This issue will be revisited in Sec. \ref{sec:pic}.
	
\begin{figure}[b]
	\includegraphics[width=8cm]{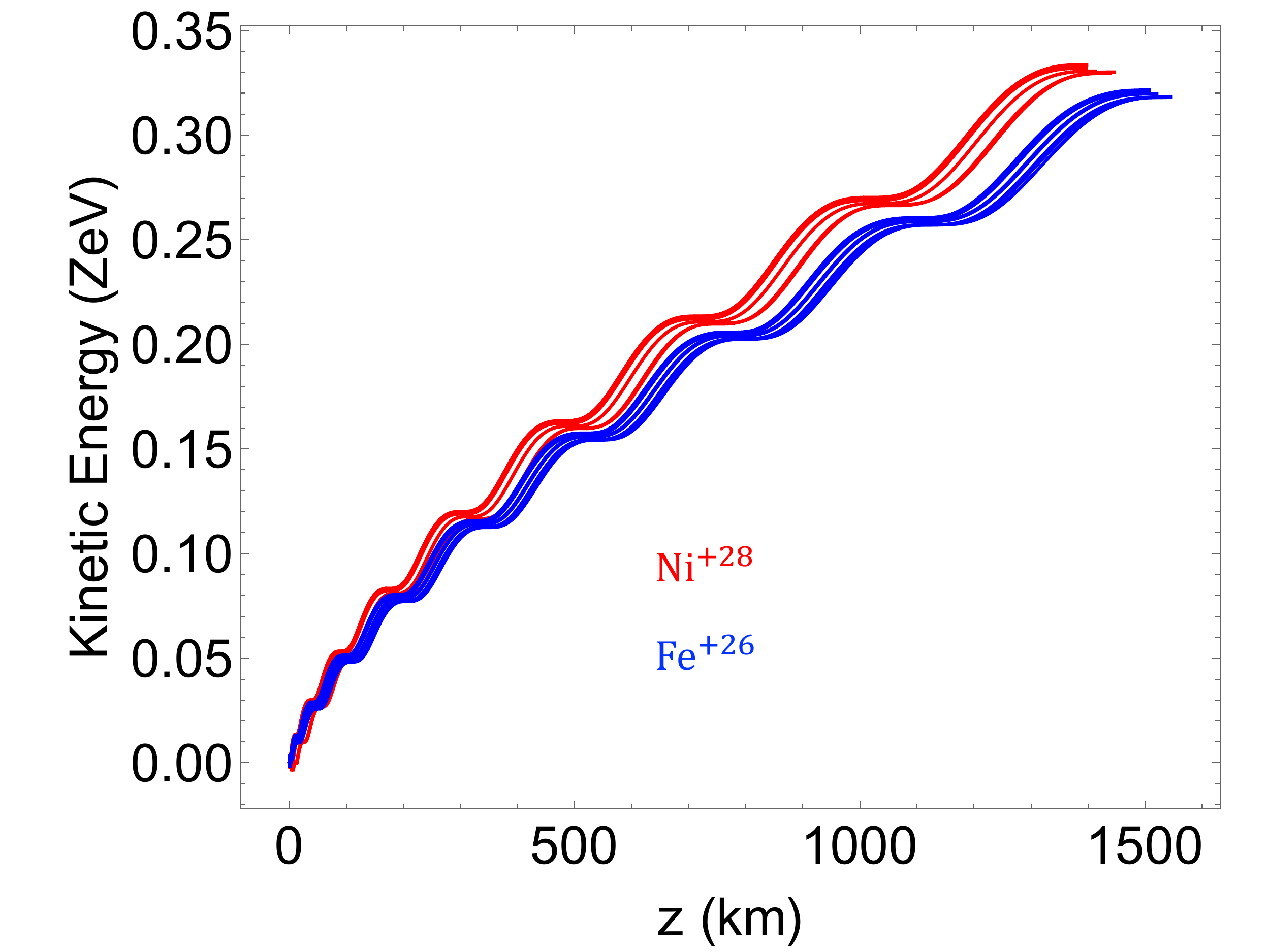}
	\caption{Kinetic energy evolution with the excursion distance of an ensemble of nuclei of Fe$^{+26}$ (blue) and Ni$^{+28}$ (red) employing the fields of infrared light of wavelength $\lambda = 0.12$ mm and intensity $I=6\times10^{32}$~ W/m$^2$. The initial ensemble has $N = 10$ nuclei inside a disk of radius 5 m and thickness $H=\lambda$ (number density $n_d \simeq 1061$ m$^{-3}$). Initial ensemble kinetic energy: normal distribution of mean $\bar{K}_0=1$ GeV and standard deviation $\Delta K_0=10$ MeV. The average resonance magnetic field strengths sensed by the ensemble members and calculated on the basis of Eq. (\ref{r}) are $B_s \simeq 287747\pm 202$ T (iron) and $298794\pm161$ T (nickel). Interaction time is equivalent to 5 phase cycles ($\Delta\eta=10\pi$).}
\label{fig5}
\end{figure}

Our aim thus far has been to lend support to the single-particle results \cite{Salamin_2021} by performing simulations for non-interacting many particles. Not only do the many-particle simulations agree, in general, with the single-particle calculations, but they do not seem to depend on the size of the ensemble employed. 

These points will be addressed together by employing a more realistic parameter set. Key departure from the old parameter set is employing infrared radiation of wavelength 0.12 mm and intensity $I=6\times10^{32}$~ W/m$^2$. An ensemble of only 10 nuclei is employed, making the particle density $n_d\simeq1061$ m$^{-3}$ and, therefore, strengthening the argument in support of the particle-particle interactions being negligibly small. Simulations have been performed for Fe$^{+26}$ and Ni$^{+28}$, two of the most stable nuclei in nature \cite{adriani}.
	
Figure \ref{fig5} shows evolution of the kinetic energies of the particles of ensembles of nuclei of iron and nickel during interaction with 5 phase cycles of the radiation field. All other ensemble parameters are given in the figure caption. The exit kinetic energies are as follows:  $K_e\simeq0.3197\pm0.0014$ ZeV, for iron, and $K_e\simeq0.3317\pm0.0013$ ZeV, for nickel. Note that the difference between the two sets of results is quite small. This is due to the fact that the masses and charge-to-mass ratios of the two nuclei are quite close. Also, the resonance magnetic field values are comparable.

\section{one-dimensional Particle-in-cell simulations}\label{sec:pic}

To approximately asses the effects associated with the particle-particle interactions in CARA, we consider injection initially of a high-density ion slab into the interaction region with the radiation and magnetic fields under near-resonance conditions. The above investigations have shown that the acceleration length (axial excursion length of the accelerated particles) can be extremely large, essentially many many kilometers. For that reason alone, PIC simulations cannot be employed. As will be seen below, dephasing and spatial spreading will quickly result in a substantial drop in the particle density. Thus, the particle-particle interaction effects on the exit kinetic energies of the accelerated particles are expected to be negligibly small. 

Despite the above, one-dimensional PIC simulations have been performed using EPOCH \cite{Arber2015} over a small part of the axial excursion length. In each simulation,
a 160 $\mu$m simulation segment, covering the range $0<z<160~ \mu$m, moves along the $z-$axis at the speed of light. The segment is subdivided into $160\times512$
cells with 64 particles per cell. The interaction scenario involves He$^{+2}$ ions initialized into a 1-$\mu$m slab with a density of $n_{i0}=10^{18}$ cm$^{-3}$, energy $K_0=150\pm1.5$ MeV, and a longitudinal static magnetic field $\vec{B}_0=B_0\hat{z}$, with $B_0=29.65$ MT obtained from the resonance condition. The plane electromagnetic (PEM) wave is initialized with a wavelength of $\lambda_0=1 ~\mu$m, intensity of $I_0=10^{30}$ Wm$^{-2}$, and a longitudinally super-Gaussian profile $f(t)=\exp{[-(t-\tau_0)^6/\tau_l^6]}$, where $\tau_0=55~ T_0$ is pulse center, and $\tau_l=110~ T_0$ is pulse width (with $T_0=\lambda_0/c$). In the interaction scenario, the ion slab is first injected into the region $0<z<1~ \mu$m of the simulation segment. Then the PEM is launched behind the slab and overtakes it.

\begin{figure*}
\includegraphics[width=18cm]{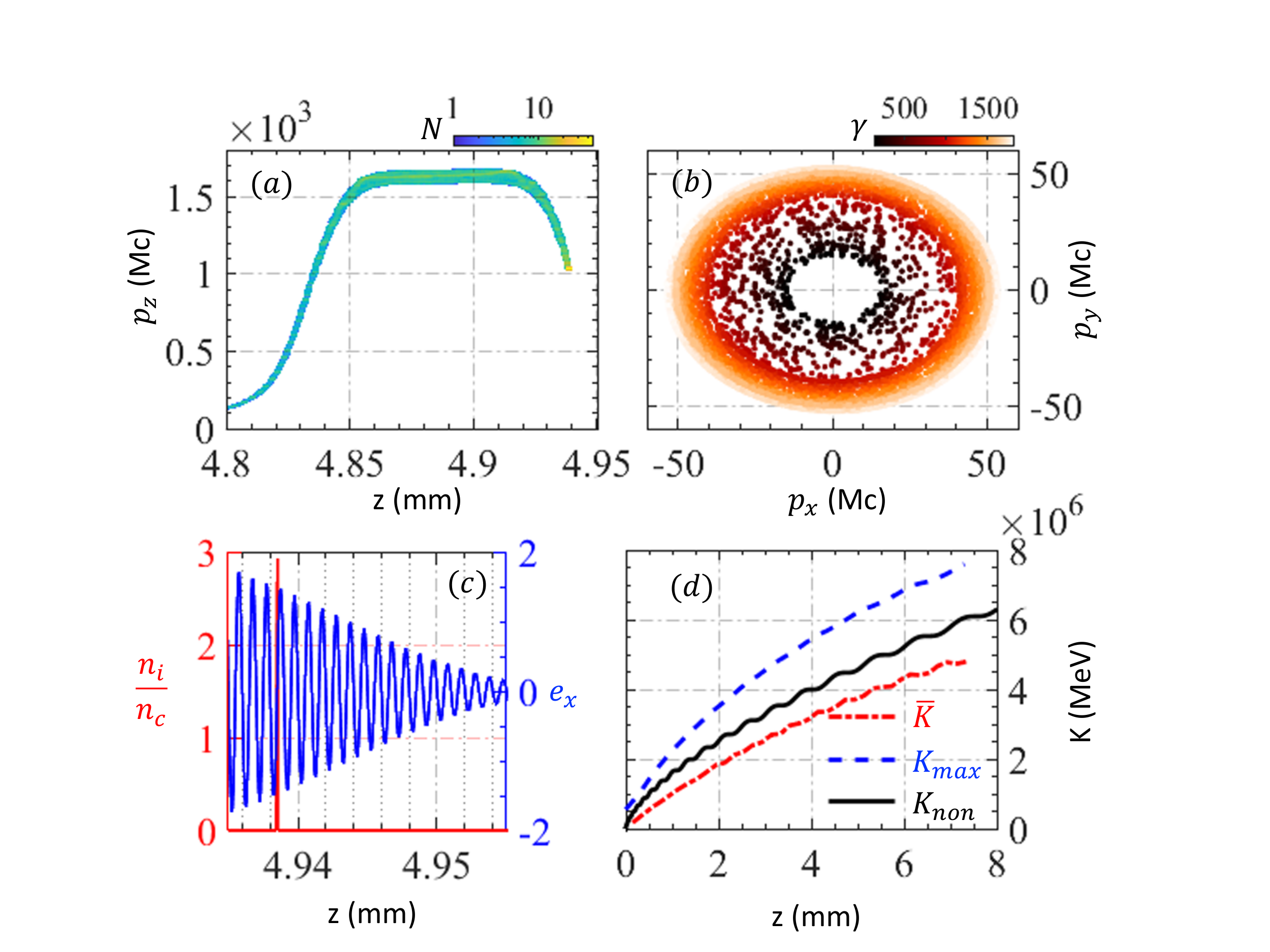}
\caption{(a)-(c) Snapshots, at propagation time $t=5000~ T_0$, for: (a) Distribution of the particle number $N$ in phase space (color bar in logarithmic scale), (b) Distribution of particle energy in the transverse momentum space, and (c) The local (sensed by the ions in the slab) electric field component, $e_x$, scaled by $E_0$ of the PEM wave. Also shown in (c) is the particle number density $n_i$, normalized by the critical density $n_c=1.116\times10^{21}$ cm$^{-3}$. All quantities displayed are for $\alpha-$particles (He$^{+2}$ ions) in interaction with a PEM wave of 1 $\mu$m wavelength and a magnetic field of strength $B_0=29.65$ MT. (d) Evolution of the energy gains with axial excursion distance. Here, $\bar{K}$ is the average kinetic energy  of the phase-locked particles inside the particle slab,  $K_{non}$ is corresponding quantity in the non-interacting case based on Eq. (\ref{gammares}), $K_{max}$ is the maximum kinetic energy of the dephased particles.}
\label{fig6}
\end{figure*}

During the propagation, because of the particle-particle Coulomb interactions and interaction of the particles with the current-induced magnetic field, in addition to the energy spread of the initial ensemble, many ions get disturbed continuously and violate the resonance condition, i.e., they phase-slip or get dephased. The dephased ions are thus knocked out of the slab and slip back through the flat top and down the ramp of the PEM pulse. Consequently, they form the belt distribution in phase space shown in Fig. \ref{fig6}(a). As may be inferred from the phase-space trajectory oscillations, slipping back through the flat top causes the dephased ions to gain higher energies, while going down the PEM ramp results in energy loss. From the energy distribution in the transverse momentum space shown in \ref{fig6}(b) one can see that the rise in energy gain is associated with increasing transverse momentum. Note also that the maximum transverse momentum is a constant, determined by the specific set of values taken by the parameters $I_0$, $B_0$ and $K_0$.

Propagation of the ion slab is illustrated in Fig. 6(c). It is shown that the front-edge phase of the PEM catches up with the slab and co-propagates with it. Density of the ion slab is increased by an order-of-magnitude compared to the initial density due to the compression caused by the focusing current-induced magnetic field. On the other hand, the slab is shortened due to the loss of dephased ions.

The gain curves, showing the maximum kinetic energy $K_{max}$ of the dephased ions and the averaged kinetic energy $\bar{K}$ of the phase-locked ions within the slab, are presented in Fig. \ref{fig6}(d), together with the average energy gain of the non-interacting particles $K_{non}$. Note that $\bar{K}$ of the ion slab behaves roughly the same as $K_{non}$ obtained from the many- and single-particle calculations based on Eq. (\ref{gammares}) with constant $a_0=1.5$, corresponding to $a_x\equiv Qe_x/(M\omega c)\simeq1.5$, where $e_x$ is the $x-$component of the local electric field sensed by the ions in the slab. The average kinetic energy $\bar{K}$ is in general lower than $K_{non}$ due to the effect of up-ramp of the PEM pulse. It seems from Fig. \ref{fig6}(d) that, after propagation for 8 mm, the particle-particle interaction effects lower the energy gain from about 6.3 to 5 TeV (or by roughly 20\%) and the particle density by more than an order-of-magnitude.

\section{Concluding remarks}\label{sec:conc}
	
This work is part of efforts to lend support to conclusions arrived at recently regarding the acceleration to ZeV energies, of protons and other bare atomic nuclei, by cyclotron autoresonance, in astrophysical environments such as the merger of a binary neutron-star system or a magnetar-powered supernova explosion. Those conclusions were based on single-particle calculations \cite{Salamin_2021}, as well as non-interacting many-particle simulations \cite{salaminPLA} in which the number density is kept well below that of a typical solid. The current study has advanced the calculations to the level of interacting many particles, employing a particle number density close to that of a solid. The process is assumed to take place outside a compact object, or equivalent, so that the particle-particle interactions may be ignored. That the particle-particle interaction effects can be ignored has been supported by PIC simulations employing, initially, a thin high-density ion slab. The effects can be strongest between the ions initially. However, interaction with the combined radiation and magnetic fields, as well as the inter-particle Coulomb forces and interaction with the ion-generated currents, change that quickly. Many ions violate the resonance condition, get dephased and slip back and to the sides. This causes the (accelerated) ion density to fall sharply and the particle-particle separations to grow axially and render the particle-particle interactions negligibly small. However, these conclusions cannot be easily generalized to apply to the ions over many kilometers.
	
Results of the many-particle simulations strongly agree with those of the single-particle calculations, with the recognition that the former is more statistically significant. Let $X$ stand for a physical quantity pertaining to the particles accelerated by CARA. In \cite{Salamin_2021} and in the present study $X\in\{K, x, y, z, B_s\}$. Denote by $X_e$ and $X'_e$ the exit values of $X$ obtained from the single-particle calculations and the many-particle simulations, respectively. In all cases considered, and for all quantities investigated, $X_e$ is found to lie within the range $\bar{X'}_e\pm\Delta X'_e$, where $\bar{X'}_e$ and $\Delta X'_e$ represent the mean and standard deviation of $X'_e$, respectively. Finally, inspection of the numbers (for the protons, in particular) displayed in Table \ref{tab:tab1}, on the one hand, and \ref{tab:tab2} and \ref{tab:tab3}, on the other, reveals that the end results do not depend strongly on the size of the ensemble.\\

\begin{acknowledgments}
This work is supported in part by the National Natural Science Foundation of China (Grant number 12105217). The work of YIS has been funded by an Alexander von Humboldt Fellowship (Wiederaufnahme) and a Faculty Research Grant (FRG number AS1811) from the American University of Sharjah.
\end{acknowledgments}

\bibliographystyle{apsrev4-2}
\bibliography{Zevatron}

\end{document}